\documentstyle[12pt]{article}

\def\beq{\begin{equation}}
\def\eeq{\end{equation}}
\def\bea{\begin{eqnarray}}
\def\eea{\end{eqnarray}}

\def\ba{\begin{array}}
\def\ea{\end{array}}
\def\bce{\begin{center}}
\def\ece{\end{center}}

\begin{document}
\begin{titlepage}
\rightline{ BROWN-HET-1148}
\rightline{hep-th/9810244}
\def\today{\ifcase\month\or
January\or February\or March\or April\or May\or June\or
ppJuly\or August\or September\or October\or November\or December\fi,
\number\year}
\vskip 1cm
\centerline{\Large \bf Three Dimensional SCFT from M2 Branes at}
\vskip .2cm
\centerline{\Large\bf  Conifold Singularities}
\vskip 1cm
\centerline{\sc  Kyungho Oh$^{a}$ and Radu Tatar$^{b}$}
\vskip 1cm
\centerline{ $^a$ Dept. of Mathematics, University of 
Missouri-St. Louis,
St. Louis, MO 63121, USA }
\centerline{{\tt oh\@arch.umsl.edu}}
\centerline{$^b$ Dept. of Physics, Brown University,
Providence, RI 02912, USA}
\centerline{\tt tatar\@het.brown.edu}
\vskip 2in
\centerline{Abstract}
Recently it was conjectured that parallel branes at
conical singularities
are related to string/M  theory on $AdS \times X$ where $X$ 
is an
Einstein manifold. In this paper we consider coincident M2
branes near a
conifold singularity when M theory is compactified on
$AdS_4 \times Q^{1,1,1}$ for $Q^{1,1,1} = (SU(2) \times
SU(2) \times
SU(2))/(U(1) \times U(1))$ as a seven dimensional
Sasaki-Einstein
manifold.
We argue that M theory on $AdS_4 \times Q^{1,1,1}$ can be 
described
in terms of a three dimensional superconformal field
theory.
We use the fact that the
three dimensional self-mirror duality
is preserved by exact marginal operators, as
observed by Strassler.
\end{titlepage}
\newpage
\section{Introduction}
\setcounter{equation}{0}
Recently, Maldacena argued that type IIB string theory on
$AdS_5 \times {\bf S^5}$ is equivalent to ${\cal N} = 4$ supersymmetric SU(N)
gauge theory in four dimensions \cite{mal} in the 't Hooft large N limit
. The correspondence was made more precise in \cite{gkp,wit1} as
stating a one to one correspondence between the Green's functions in type
IIB string theory on $AdS_5 \times {\bf S^5}$ and the correlation
functions
of gauge invariant operators of the ${\cal N} = 4$ supersymmetric
field theory. 

Two other conjectures were made in \cite{mal} by relating M theory
compactified on $AdS_4 \times {\bf S^7}$
or $AdS_7 \times {\bf S^4}$ and ${\cal N} = 8$ three dimensional field
theory
or (0, 2) six dimensional field theory. The correspondence between the Green's
function in M theory and the correlation function of the field theory
has been studied in \cite{aoy,lr,min,hal}. 
Maldacena's conjectures have been generalized to theories with less
supersymmetries by orbifolding \cite{ks,lnv,ot,gom1,fkpz,be,eg,aot1,aot2}
and to F theory \cite{fs,afm,aot3}.

One interesting generalization is to consider five-dimensional
manifolds other than ${\bf S^5}$ \cite{kw, mp}. (See also \cite{gk,afhs,f1}
for important discussions over these manifolds.) The fact that ${\bf S^5}$
preserves the maximum amount of supersymmetry while  Einstein
manifolds does not in general implies  that the field theory
obtained in this way have fewer supersymmetry. In \cite{kw} Klebanov 
and Witten 
identified the field theory coming from string theory compactified on
$AdS_5 \times X_5$ where $X_5$ was taken to be the homogeneous space
$T^{1,1} = (SU(2) \times SU(2))/U(1)$. The theory on the worldvolume of D3 
branes is ${\cal N} = 1$
superconformal field theory in four dimensions. 
It was also suggested the possibility to extend the result to M theory.
In \cite{mp}, Morrison and  Plesser investigated M or IIB theory compactified on $AdS_{p+2} \times H^{D-p-2}$ where $H$ are Einstein manifolds obtained as 
the horizons  of the  
Gorenstein canonical singularities. 

In a similar fashion to \cite{kw}, we are going to consider a specific
geometry $AdS_4 \times X^7$ where $X_7$ is a seven-dimensional compact
Einstein manifold given by 
$Q^{1,1,1} = (SU(2) \times SU(2) \times SU(2))/(U(1) \times U(1))$.   
Historically, the nomenclature $Q^{p, q, r}$ manifolds is due to the
classification of all Kaluza - Klein coset G/H seven-manifolds
performed by Castellani, Romans and Warner in \cite{crw}. This
classification both
systematized already existing construction and introduced new manifolds.
In particular the $Q^{pqr}$ manifolds had already been constructed and
their
supersymmetry derived by D'Auria, Fre' and Van Nieuwenhuizen in
\cite{afn} (see also \cite{np,pp}). The use of G/H Einstein manifolds to
generate M2--brane
solutions in
connection with Conformal Field Theory interpretation has been discussed
recently by Cersole et al. in \cite{cere}. 

We will present  a Gorenstein canonical singularity $Y$
whose horizon manifold
is  $Q^{1,1,1}$. Hence $Y$ can be regarded as a limiting space of 
Calabi-Yau manifolds.  
We will also find interesting
relations with theories with M2 branes placed at an orbifold singularity
${\bf S^5/Z_2 \times Z_2}$. 
The homogeneous space $Q^{1,1,1} = (SU(2) \times SU(2) \times SU(2))/(U(1)
\times U(1))$ may be obtained by blowing-up the orbifold 
${\bf S^5/Z_2 \times Z_2}$ which can be seen in type IIA as
D2 branes in the presence of 2 D6 branes at 
${\bf R^4/Z_2} \times{\bf R^4/Z_2}$ singularity or as
an elliptic model with D3 branes on a circle together with 2 NS5 branes and
2 D5 branes. This model is self-mirror and we use its self-duality to
argue the existence of an exact marginal operator which gives a
superconformal field theory.

In section 2 we begin with a brief introduction to Sasaki-Einstein
manifolds and provide a Calabi-Yau Gorenstein canonical singularity $Y$
whose horizon is the Sasaki-Einstein manifold $Q^{111}$.
In section 3 we show that topologically the horizon $Q^{1,1,1}$ is the same as
the complex blow-up of the orbifold 
 ${\bf S^5/Z_2 \times Z_2}$ as two ${\bf S^3}$
fibrations over ${\bf S^2 \times S^2}$, which leads us to  
construct a field theory inspired by
a theory on the complex blow-up of the orbifold ${\bf S^5/Z_2 \times Z_2}$. 
In section 4 we discuss a field
theory obtained on the brane world-volume placed on the Calabi-Yau
Gorenstein canonical
singularity $Y$. The field theory discussion was motivated  by the recent work
of Strassler~\cite{ms}.
\section{Near-Horizon Geometries of Cone Branes}
\setcounter{equation}{0}

Let $(X, g_{X})$ be a Riemannian manifold of real dimension $2n-1$
and ${\bf R}_+$ be
the open half-line $0< r < \infty$. Then the metric cone C(X)
over $X$ is  a Riemannian manifold 
${\bf R}_+ \times X$ with a metric 
\bea
g_{C(X)} = dr^2 + r^2 g_X
\eea
We often add one point (which is called the vertex) to $C(X)$
corresponding to the location  $r =0$ and 
use the same notation $C(X)$ when there is no danger of confusion.

>From a relation between Ricci curvature of $X$ and its  metric cone  $C(X)$,
we can show that  $X$ is Einstein with positive curvature if and only if 
then $C(X)$ is Ricci-flat~\cite{fk, kw}. However $C(X)$ will be
metrically singular at the vertex of the cone except
the sphere, in which case  $C(X)$ is actually flat.

We say that $(X,g_X)$ is {\it Sasakian} if the holonomy 
group of the metric cone $(C(X), g_{C(X)})$  reduces to a subgroup of
$U(n)$. In particular, $(C(X), g_{C(X)})$ is K\"ahler.
Therefore, $(X,g_X)$ is Sasaki-Einstein if and only if its metric cone
$C(X)$ is Calabi-Yau (K\"ahler Ricci-flat).

Let $I$ be a parallel complex structure on $C(X)$, i.e. $I$ commutes with
the Levi-Civita connection on $C(X)$. Then $\chi := I(\partial r)$ will  be 
a unit Killing vector field on $X$ where $X$ is identified 
with $X \times \{1\}$. The Killing vector field $\chi$ on $X$ defines a 
foliation ${\cal F}$ whose leaves are the integral curves of $\chi$. 
$X$ is called a {\it  regular Sasakian} manifold when these leaves are 
closed and have the same length. In this case, $\chi$
 defines a $U(1)$ action on $X$
and $X$ can be understood as a circle bundle over the orbit 
space $(2d-2)$ real dimensional manifold $M$, which is the space of the leaves.
The CR structure on $X$ pushes down to give a complex structure on $M$ and
the Sasakian condition on $X$ will guarantee that the complex structure will
be K\"ahler. By a version of the Kodaira Embedding Theorem~\cite{ba},
$M$ will be a projective variety. Thus $C(X)$ is a ${\bf C}^*$-bundle over
a projective variety $M$ and this can be realized as an affine cone
over $M$ (which will be defined for a Sasaki-Einstein 7-manifold
$Q^{1,1,1}$
later).
It also can be shown that a ${\bf Q}$-factorial Fano variety. For details, we  refer the reader to ~\cite{bg}.
\par
The above process can be reversed in the following sense.
Consider a complex variety $Y \subset {\bf C}^n$ 
of complex dimension $d$ such that $0\in Y$. Let $\rho:{\bf C}^n \to {\bf R}$
be the square of the usual distance function, namely,
\bea
\rho (y) = |y_1|^2 + \cdots + |y_n|^2
\eea
for $y = (y_1, \ldots , y_n) \in  {\bf C}^n$. It is easy to see that
there exists an $\epsilon_0 > 0$ such that all the $(2d-1)$ real dimensional
spheres 
$S_\epsilon = \rho^{-1}(\epsilon )$ for $\epsilon_0 \geq \epsilon > 0$
are transversal to $Y$~\cite{gwpl}. In particular, $S_\epsilon \cap Y$ 
is  smooth  and has the equivalent Riemannian structure
for
all $\epsilon \in (0, \epsilon_0)$. We denote it by $L(Y,0)$ and call it 
the {\it link} (or horizon) of $Y$ at $0$.
In fact, integration along
the Euler vector fields 
$\rho \partial / \partial \rho$ produces an isometry
\bea
\phi: L(Y,0) \times (0, \epsilon_0) \to 
(B_{\epsilon_0} \setminus \{ 0\}) \cap Y
\eea
such that $\rho(\phi (x,t)) = t$ where 
$B_{\epsilon_0} = \{ x \in {\bf C}^n : \rho(x) < \epsilon_0 \}$.
Hence $(B_{\epsilon_0} \setminus \{ 0\}) \cap Y$ is the metric cone over the link
$ L(Y,0)$. The link (or horizon) $ L(Y,0)$ is  Sasaki-Einstein if and only if 
the singularity $(Y, 0)$ is Calabi-Yau. Let $I$ be the complex structure on
$Y$. If $ L(Y,0)$ is regular, then the Killing vector field 
$\chi :=I(\rho \frac{\partial }{\partial \rho} )$ defines a foliation
whose leaves are circles of the same length. This implies that the ${\bf C}^*$-action on ${\bf C}^n$ induces an action on $Y$ and $M$ is the quotient. Thus
$M$ is a projective variety and 
$Y$ is an affine cone over $M$.
\par
In this paper, we are interested in a regular Sasaki-Einstein 7-manifold
$Q^{1,1,1}$ which is known to be a $U(1)$ bundle over a threefold 
${\bf CP}^1 \times {\bf CP}^1 \times {\bf CP}^1$ with winding number $(1,1,1)$.
Recall~\cite{afn} that $Q^{p,q,r}$ is a homogeneous space 
$(SU(2) \times SU(2) \times SU(2))/(U(1) \times U(1))$ where $U(1) \times U(1)$
is embedded as follows: 
Consider an embedding of $U(1)\times U(1) \times U(1)$ as the standard
maximal torus of $(SU(2) \times SU(2) \times SU(2))$. Thus each $U(1)$  
is embedded into $SU(2)$ via the third Pauli matrix on the Lie algebra level.
Let $\bf{h}= {\bf u(1)}\oplus {\bf u(1)}$ be
the Lie subalgebra orthogonal to   the Lie subalgebra  generated by
$p\sigma_{z,1} + q \sigma_{z,2} +r \sigma_{z,3}$ in the Lie algebra
${\bf u(1)}\oplus {\bf u(1)}\oplus {\bf u(1)}$ 
where $ \sigma_{z,i}$ are the generators of  ${\bf u(1)}$.
Then we embed  $\bf{h}$ into ${\bf{su(2)}}\oplus {\bf{su(2)}}\oplus {\bf{su(2)}}$ by the embedding described above. Its quotient
in the Lie group level is denoted by $Q^{p,q,r}$. By taking further 
quotient of $Q^{p,q,r}$ by the Lie subalgebra generated by
$p\sigma_{z,1} + q \sigma_{z,2} +r \sigma_{z,3}$,
one can see that $Q^{p,q,r}$ is a $U(1)$-bundle over 
$(SU(2)/U(1)) \times (SU(2)/U(1)) \times (SU(2))/U(1))$ which is 
${\bf CP}^1\times {\bf CP}^1\times {\bf CP}^1$. The complexification of
this $U(1)$-bundle over ${\bf CP}^1\times {\bf CP}^1\times {\bf CP}^1$
will be ${\cal O}_{{\bf CP}^1}(-p) \otimes {\cal O}_{{\bf CP}^1}(-q) 
\otimes {\cal O}_{{\bf CP}^1}(-r)$. If we embed  
${\bf CP}^1\times {\bf CP}^1\times {\bf CP}^1$ by 
${\cal O}_{{\bf CP}^1}(p) \otimes {\cal O}_{{\bf CP}^1}(q) 
\otimes {\cal O}_{{\bf CP}^1}(r)$ into 
${\bf P}(H^0({\cal O}_{{\bf CP}^1}(p) \otimes {\cal O}_{{\bf CP}^1}(q) 
\otimes {\cal O}_{{\bf CP}^1}(r)))$ and take the affine cone over the
image, the vertex will be singular and the exceptional divisor over
the vertex in the blowing-up will be ${\cal O}_{{\bf CP}^1}(-p) \otimes 
{\cal O}_{{\bf CP}^1}(-q) 
\otimes {\cal O}_{{\bf CP}^1}(-r)$. 

We will provide more details for $Q^{1,1,1}$.
Consider the following  embedding 
\bea
\sigma :{\bf CP}^1 \times {\bf CP}^1 \times {\bf CP}^1 \hookrightarrow
{\bf CP}^7
\eea
 by an very ample line bundle 
${\cal O}_{{\bf CP}^1 \times {\bf CP}^1 \times {\bf CP}^1}(1,1,1)$, 
which can be expressed as a map sending 
\bea
\label{su2}
(s: t) \times (u: v) \times (w: x) \longrightarrow 
(suw: sux: svw: svx: tuw: tux: tvw: tvx).
\eea
in terms of the tri-homogeneous coordinates of  
${\bf CP}^1 \times {\bf CP}^1 \times {\bf CP}^1$ and the
homogeneous coordinates of  ${\bf CP}^7$. We denote the image by
\bea
Z := \sigma ({\bf CP}^1 \times {\bf CP}^1 \times {\bf CP}^1)
\eea
Let $q: {\bf C}^8 \setminus \{ 0\} \to {\bf CP}^7$ be the natural quotient
map of ${\bf C}^8 \setminus \{ 0\}$ by the ${\bf C}^*$-action.
Then the affine cone over 
$Z$ 
\bea
Y := q^{-1}(Z)\cup \{ 0\}
\eea
is the desired singularity of a  Calabi-Yau fourfold.  
Thus $Y$ in ${\bf C}^8$ is a zero locus of the ideal $I$ generated by
the kernel of
the map
\bea
\label{sigma}
\sigma^* : {\bf C}[z_0, z_1, \ldots , z_7] \longrightarrow
{\bf C}[s, t, u, v, w, x],\\ \nonumber
\sigma^*(z_0) = suw, \sigma^*(z_1)= sux, \sigma^*(z_2)= svw, \sigma^*(z_3)=svx,
\\ \nonumber
\sigma^*(z_4)=tuw, \sigma^*(z_5)=tux, 
\sigma^*(z_6)=tvw, \sigma^*(z_7)=tvx.
\eea
One can compute the ideal $I$ (for example, using Gr\"obner basis)
 and hence one can show
that $Y$ is  defined by the equations
\bea
\label{su3}
z_0z_3 - z_1z_2 = z_0z_5 - z_1z_4 =z_0z_6 - z_2z_4& 
 = & 0,\\ \nonumber
z_0z_7 - z_1z_6 = z_0z_7 - z_2z_5=z_0z_7 - z_3z_4
& = & 0,\\ \nonumber
z_1z_7 - z_3z_5 = z_2z_7 - z_3z_6  = z_4z_7 - z_5z_6 & = & 0.
\eea
One can see that the ideal $I$ can be generated by the quadrics from the
following minimal free resolution of ${\cal O}_Z$:
\bea
0 \to {\cal O}_{{\bf P}^7}(-6)
&\!\to\! & \bigoplus_{9}{\cal O}_{{\bf P}^7}(-4)\to 
\bigoplus_{16}{\cal O}_{{\bf P}^7}(-3)\to 
\bigoplus_{9}{\cal O}_{{\bf P}^7}(-2)\nonumber\\
\phantom{1234567}&\!\to\!& {\cal O}_{{\bf P}^7}\to {\cal O}_Z \to 0.
\label{min-res}
\eea
It is easy to check all possible quadric generators for the ideal $I$.
Now we want to show that $Y$ has a Gorenstein canonical singularity. This is 
needed to satisfy 
a requirement that $Y$ be at finite distance with respect to  the natural 
Weil-Peterson metric on the moduli of Calabi-Yau
manifolds as the limiting space in it~\cite{mp,ha,wang}.
Since $Z$ is projectively normal, $Y$ is normal. Thus $Y$ is toric because
there is a $U(1)^4$-action on $Y$. (You may also explicitly compute
the toric fan $\Delta$ which realizes this toric embedding. The fan $\Delta$ is
nothing but  a dual cone $\sigma$
in ${\bf Z}^6$ of the convex rational polyhedral
cone $\check{\sigma}$
generated by $(1,0,1,0,1,0), (1,0,1,0,0,1),$ $ (1,0,0,1,1,0), (1,0,0,1,0,1),$
 $(0,1,1,0,1,0), (0,1,1,0,0,1), (0,1,0,1,1,0), (0,1,0,1,0,1)$ corresponding
to the monomials which appears in $(\ref{sigma})$. We refer
to \cite{agm, mp2} for a review of toric geometry.) Since the vertex of $Y$ has
a rational singularity (again because it is toric), $Y$ has 
a canonical singularity when it is Gorenstein~\cite{mr}. One can prove that
$Y$ is Gorenstein by computing the dualizing sheaf $\omega_Y$
of $Y$ (which is the same as the bundle of holomorphic 4-forms on the
smooth part of $Y$) using the fact
that $Z$ is subcanonical. Note that since $Y$ is an affine toric variety,
the dualizing sheaf $\omega_Y$ is in fact 
trivial which means that there is a non-vanishing 
holomorphic 4-form on $Y$. (You can also show this fact by a direct computation
of $\omega_Y$ using the resolution (\ref{min-res}).)
As it is stated in \cite{mp}, this means that there exists a non-vanishing holomorphic 
$4$-form on $Y$ which 
extends to a holomorphic $4$-form on any smooth resolution of $Y$.

In this setting, $Q^{1,1,1} = L(Y,0)$. Thus $Q^{1,1,1}$ can be taken to be 
the intersection of Y with the unit sphere in ${\bf C}^8$:
\begin{equation}
|z_0|^2 + |z_1|^2 + |z_2|^2 + |z_3|^2 + |z_4|^2 + |z_5|^2 + |z_6|^2 +
|z_7|^2 = 1.
\label{su4}
\end{equation}

>From the way the equations for $Y$ are derived, it is clear that
the equations for $Y$ can be `solved' by setting
\bea
z_0 = S C_1 D_1, z_1 = S C_1 D_2, z_2 = S C_2 D_1, z_3 = S C_2 D_2,\\
\nonumber
z_4 = T C_1 D_1, z_5 = T C_1 D_2, z_6 = T C_2 D_1, z_7 = T C_2 D_2.
\label{su5}
\eea
where $(S, T)$, $(C_1, C_2)$, $(D_1, D_2)$ are the pairs of homogeneous
coordinates for the three ${\bf CP}^1$ spaces. 
It is easy to check that this choice for $z_i$ satisfies equations
(\ref{su3}). 

If we write now $S = A_2 B_1$ and $T = A_1 B_2$, they are invariant under
\bea
\label{u11}
(A_1, A_2)\rightarrow \lambda (A_1, A_2)\phantom{QWE}
(B_1, B_2) \rightarrow \lambda^{-1} (B_1, B_2)
\eea
 The $z_i$ are invariant under
\bea
\label{u12}
(C_1, C_2) \rightarrow \gamma (C_1, C_2) \phantom{QWE}
(D_1, D_2)  \rightarrow \gamma^{-1} (D_1, D_2) 
\eea
$\lambda$ and $\gamma$ are two complex numbers. 

We can now choose the 
real parts of $\lambda$ and $\gamma$ to set:
\begin{equation}
|B_1|^2 + |C_1|^2 + |C_2|^2 = |A_1|^2 + |D_1|^2 + |D_2|^2
\label{1}
\end{equation}
and
\begin{equation}
\label{2}
|B_2|^2  + |C_1|^2 + |C_2|^2 = |A_2|^2 + |D_1|^2 + |D_2|^2
\end{equation}

To identify the manifold $Q^{1,1,1}$, we set
$|C_1|^2 + |C_2|^2 = |D_1|^2 + |D_2|^2 = |S|^2 + |T|^2 = 1$
so the real isometry group is 
 $SU(2) \times SU(2) \times SU(2)$ and this is to be
divided
by the angular parts of (\ref{u11}) and (\ref{u12}) which give 2 U(1)
groups so finally we obtain $Q^{1,1,1} = (SU(2)\times SU(2) \times SU(2))/
(U(1) \times U(1))$.

Our goal is to find the ${\cal N} = 2$ superconformal theory which is dual
to the M theory compactified on $AdS_4 \times Q^{1,1,1}$, seen as the infrared
limit of the theory of N coincident M2 branes placed at a
conifold singularity of $M_3 \times Y$.

\section{Comparison to  an ${\bf R}^4/ {\bf Z}_2 \times
 {\bf R}^4/ {\bf Z}_2$ orbifold.}

In order to perform an important check over our theory, we compare the 
conifold to a ${\bf R}^4/ {\bf Z}_2 \times
 {\bf R}^4/ {\bf Z}_2$ orbifold background.
Consider an action of ${\bf Z}_2 \times {\bf Z}_2$ on ${\bf R}^8$ as follows: 
\bea
(1,0) \cdot (x_1, \ldots , x_8) = 
(-x_1, -x_2, -x_3, -x_4, x_5, x_6,x_7, x_8) \\
(0,1) \cdot (x_1, \ldots , x_8) = 
(x_1, x_2, x_3, x_4, -x_5, -x_6,-x_7, -x_8)
\eea
where $(1,0)$ (resp. $(0,1)$) is the generator of the first (resp. second)
factor of the group ${\bf Z}_2 \times {\bf Z}_2$ and $x_1, \ldots , x_n$
are the coordinates of ${\bf R}^8$.
This will induce an action on ${\bf S}^7$ in ${\bf R}^8$
given by an equation $ x_1^2 + x_2^2 + \cdots + x_7^2 = 1$.
The twisted sector mode of string theory on 
$AdS_3 \times {\bf S}^7/{\bf Z}_2 \times {\bf Z}_2$ is the blowup of the
orbifold
singularity of ${\bf S}^7/{\bf Z}_2 \times {\bf Z}_2$.

In order to understand the geometry of this blowup, we study the
blowup 
of an orbifold singularity ${\bf R}^8/ {\bf Z}_2 \times {\bf Z}_2$
in the complex sense via identification of 
${\bf C}^4$ with ${\bf R}^8$.
First note that 
\bea
\label{cong}
{\bf R}^8/ {\bf Z}_2 \times {\bf Z}_2 \cong{\bf R}^4/ {\bf Z}_2 \times 
 {\bf R}^4/ {\bf Z}_2
\eea
where ${\bf Z}_2$ acts on ${\bf R}^4$ by $-1$.
Let 
\bea
\label{blow}
\pi: Bl({\bf R}^8/ {\bf Z}_2 \times {\bf Z}_2) \to {\bf R}^8/ {\bf Z}_2 \times {\bf Z}_2 \\
\pi' :Bl({\bf R}^4/ {\bf Z}_2 ) \to  {\bf R}^4/ {\bf Z}_2 
\eea
be 
the complex blowups of the orbifold singularities
${\bf R}^8/ {\bf Z}_2 \times {\bf Z}_2$ and 
${\bf R}^4/ {\bf Z}_2 $ respectively. 
By (\ref{cong}), we have
\bea
Bl({\bf R}^8/ {\bf Z}_2 \times {\bf Z}_2) 
\cong Bl({\bf R}^4/ {\bf Z}_2 ) \times Bl({\bf R}^4/ {\bf Z}_2 ).
\eea
The space  $Bl({\bf R}^4/ {\bf Z}_2)$ can be regarded as the total space of
a line bundle ${\cal O}_{{\bf CP}^1}(-1)$. Hence it is a complex 
line bundle over
${\bf S}^2$. Therefore there is a vector bundle map
\bea
\label{bundle}
q:Bl({\bf R}^8/ {\bf Z}_2 \times {\bf Z}_2) \longrightarrow
{\bf S}^2 \times {\bf S}^2
\eea
with fibers ${\bf R}^4$.
We define the complex blowup of the orbifold singularity 
${\bf S}^7/{\bf Z}_2 \times {\bf Z}_2$ via the map $\pi$ in $(\ref{blow})$:
\bea
Bl({\bf S}^7/{\bf Z}_2 \times {\bf Z}_2) = \pi^{-1}({\bf S}^7/{\bf Z}_2 \times {\bf Z}_2).
\eea 
Thus we have the following diagram:
\newline
\\
\noindent
\begin{picture}(300, 85)(-40,0)
\setlength{\unitlength}{0.5mm}
\put(50, 40){$Bl({\bf S}^7/{\bf Z}_2 \times {\bf Z}_2)$}
\put(160, 40){$ Bl({\bf R}^8/ {\bf Z}_2 \times {\bf Z}_2)$}
\put(60, 0) {${\bf S}^7/{\bf Z}_2 \times {\bf Z}_2$}
\put(170, 0){${\bf R}^8/ {\bf Z}_2 \times {\bf Z}_2$.}
\put(80, 35){\vector(0,-1){25}}
\put(190, 35){\vector(0,-1){25}}
\put(115, 42){\oval(6,4)[l]}
\put(115, 2){\oval(6,4)[l]}
\put(130, 42){$i_{Bl}$}
\put(130, 2){$i$}
\put(115, 40){\vector(1,0){36}}
\put(115, 0){\vector(1,0){36}}
\end{picture}
\\
Now we study the complex blowup $Bl({\bf S}^7/{\bf Z}_2 \times {\bf Z}_2)$.
There is a map 
\bea
\label{qs}
q_S: Bl({\bf S}^7/{\bf Z}_2 \times {\bf Z}_2) 
\rightarrow {\bf S}^2 \times {\bf S}^2
\eea
which is a composition of the natural inclusion $i_{Bl}$
and 
the map $q$ in $(\ref{bundle})$. 
Since $i_{Bl}$ and  $q$ are transversal, the map $q_S$  in $(\ref{qs})$
is
a smooth fibration and it is easy to see the fiber is  ${\bf S}^3$.
In a summary, we have the following fibrations:
\newline
\\
\noindent
\begin{picture}(300, 130)(-40,0)
\setlength{\unitlength}{0.5mm}
\put(50, 80){$Bl({\bf S}^7/{\bf Z}_2 \times {\bf Z}_2)$}
\put(160, 80){$ Bl({\bf R}^8/ {\bf Z}_2 \times {\bf Z}_2)$}
\put(20, 45) {${\bf S}^3$}
\put(130, 45){${\bf R}^4$}
\put(194, 40){$q$}
\put(180, 0) {${\bf S}^2 \times {\bf S}^2$}
\put(32, 57){\vector(3,2){27}}
\put(142, 57){\vector(3,2){27}}

\put(190, 75){\vector(0,-1){65}}
\put(115, 82){\oval(6,4)[l]}
\put(45, 47){\oval(6,4)[l]}
\put(130, 82){$i_{Bl}$}
\put(115, 80){\vector(1,0){36}}
\put(45, 45){\vector(1,0){66}}
\end{picture}
\\
Thus $Bl({\bf S}^7/{\bf Z}_2 \times {\bf Z}_2)$ is an ${\bf S}^3$ bundle over
${\bf S}^2 \times {\bf S}^2$.
Now we claim that both $Q^{1,1,1}$ and $Bl({\bf S}^7/{\bf Z}_2 \times {\bf Z}_2)$
are topologically trivial ${\bf S}^3$ bundles  over  
${\bf S}^2 \times {\bf S}^2$. In order to show that $Bl({\bf S}^7/{\bf Z}_2 \times {\bf Z}_2)$
is  topologically trivial ${\bf S}^3$ bundle, note that the blow-up 
$Bl({\bf S}^7/{\bf Z}_2 \times {\bf Z}_2)$ can be achieved in two steps.
First we consider a blow-up $Bl({\bf S}^7/{\bf Z}_2 )$, where ${\bf Z}_2$
acts only on the first four coordinates by $-1$. Then the situation is
the same as in \cite{kw} except we have two more extra coordinates. 
Thus it will be a trivial ${\bf S}^5$ bundle over ${\bf S}^2$. Now we further 
blow-up the space $Bl({\bf S}^7/{\bf Z}_2 )/{\bf Z}_2 $, where ${\bf Z}_2$ acts
on the last four coordinates.  This will act on the fiber  ${\bf S}^5$ 
and it will be the trivial bundle over ${\bf S}^2$.
Since the space $Bl({\bf S}^7/{\bf Z}_2 \times {\bf Z}_2)$ is the same as the
space we obtain by the success of two blow-ups we described, we can conclude
that $Bl({\bf S}^7/{\bf Z}_2 \times {\bf Z}_2)$ is topologically
${\bf S}^3 \times {\bf S}^2  \times {\bf S}^2$. In the case of 
$Q^{1,1,1} = (SU(2)\times {SU(2)}\times
 {SU(2)})/ ({U(1)} \times  U(1))$,
the triviality can be proved by observing that 
the `forgetting' map of
the last $SU(2)$ will give a fibration of ${\bf S}^3$ over 
${\bf S}^2 \times {\bf S}^2$, which is trivial.
\section{${\cal N}=2$ field theory in 3 dimensions}
\setcounter{equation}{0}
In this section we are going to build the field theory corresponding to
the M theory compactified on the conifold Y. In order to do this we 
are going to use the above identification of both $Q^{1,1,1}$ and the 
blow-up of the orbifold $Bl({\bf S}^7/{\bf Z}_2 \times {\bf Z}_2)$ as 
${\bf S}^3$ bundles  over ${\bf S}^2 \times {\bf S}^2$. 
\footnote{The field theory discussion was  suggested to us by Matthew
Strassler and we refer to \cite{ms} for a detailed and interesting
discussion about exactly marginal operators in three dimensions.}

\vskip .2cm  
  
{\it Orbifold theory}
 
\vskip .2cm

We begin by briefly reviewing  the discussion
of \cite{kw}. 
Klebanov and Witten considered
 $AdS_5 \times {\bf S^5/Z_2}$ background where
the ${\bf Z_2}$ acts on four of the six coordinates of ${\bf S^5}$. 
In terms of branes, this corresponds to D3 branes moving on a 
${\bf C^2/Z_2 \times C}$ space which after taking a T-duality on one of the
directions of ${\bf C^2/Z_2}$ gives an elliptic model with 
D4 branes between NS5 branes. The gauge group is $SU(N) \times SU(N)$, at
each NS5 branes we have an $(N,\bar{N})$ field. The rotation of NS5
branes correspond to adding a mass for the adjoint field which gives
a $N = 1$ theory in which the $(N,\bar{N})$ and $(\bar{N}, N)$
fields correspond to two $(N,\bar{N})$ fields and two $(\bar{N}, N)$
fields. There is one SU(2) group acting on the two $(N,\bar{N})$ fields
and one SU(2) group acting on the two $(\bar{N}, N)$ fields. 

We use the same argument in our case where we start with M2 branes on
${\bf C^2/Z_2 \times C^2/Z_2}$ space. \footnote{This choice for orbifold 
was suggested to us by Matthew Strassler who pointed out that by making
this choice we would obtain self mirror-symmetrical models where we could 
have a nice identification of marginal operators}
We want to go to type IIB string theory in 10 dimension. We consider this
two
$A_1$ singularity to be given by a ${\bf Z_2}$ orbifold combined with 
two D6 branes. In 11 dimensions we start with 2 types of KK monopoles, say
in the 
$(x^3, x^4, x^5, x^6)$ directions and $(x^7, x^8, x^9,x^{10})$ directions
respectively, and with M2 branes in the $(x^{1},x^{2})$ directions. 
Reduction to 10 dimensions gives KK monopoles in the $(x^3, x^4, x^5,
x^6)$ directions, D6 branes in $(x^{1}, x^{2}, x^3, x^4, x^5, x^6)$
directions and D2 branes in the $(x^{1},x^{2})$ directions.
If the $x^3$ direction is compact, we make a T - duality with respect to
it and we obtain a configuration with D3 branes on $(x^{1}, x^{2}, x^3)$
directions, D5 branes in the $(x^{1}, x^{2}, x^4, x^5, x^6)$ directions
and NS5 branes in $(x^{1}, x^{2}, x^7, x^8, x^9)$ directions. 

The configuration we obtain is an elliptic, self mirror model  
having
N D3 branes on a circle intersecting two non-coincident D5 branes and 
two non-coincident NS branes,
with the above orientation. The theory obtained is an ${\cal N} = 4$
supersymmetric field theory with a gauge 
group $U(N) \times U(N)$ and matter given
by one field $A_{1}$ in the $(\bar{N}, 1)$ representation, 
one field $B_1$
in the $(N, 1)$ representation, 1 field $A_2$ in the $(1, \bar{N})$
representation, 1 field $B_{2}$ in the $(1, N)$
representation,
2 fields $C_{1}, C_2$ in the $(N, \bar{N})$ representation and two fields
$D_1, D_2$ in
the $(\bar{N}, N)$ representation where $ i = 1, 2$ (we discuss the
fields in the ${\cal N} = 2$ language). The model also
contains fields in the adjoint representation of the two U(N), denoted by
$\Phi$ and $\tilde{\Phi}$. The $A_i, B_i$ fields are given by strings with
one end on D3 and one end on either of the two D5 branes, the $C_i, D_i$
fields are given by strings with both ends on the D3 branes stretched 
across either of the two NS branes. 
The superpotential is 
\begin{equation}
\label{super}
g (A_1 \Phi B_1 + A_2 \tilde{\Phi} B_2) +
   g  \mbox{Tr} [\Phi (C_1 D_1 + C_2 D_2)] + g \mbox{Tr} [\tilde{\Phi}
(D_1 C_1  + D_2 C_2)]
\end{equation}
 
To arrive to a ${\cal N} = 2$ theory, we add a mass term for the $\Phi$ and 
 $\tilde{\Phi}$. This corresponds to rotating the NS branes in the 
$(x^5, x^6, x^8, x^9)$ directions.  But the rotation of the
D5 brane induces a change in the coupling between 
the hyper-multiplets A and B and the adjoint multiplets. 
In order to preserve the self-duality, we
need to rotate the NS5 branes and the D5 branes by the same angle.
In other words, we need to add the following to the superpotential: 
\begin{equation}
\label{mass}
\frac{m}{2} (\mbox{Tr} (\Phi^2 + \tilde{\Phi}^2) + \mbox{Tr} (A_1 B_1 - A_2
B_2)^2
+ \mbox{Tr} (C_1 D_1 - C_2 D_2)^2)
\end{equation}
In equation (\ref{mass}),  $ m$ is the mass of the adjoint field and the 
tangent of the rotation angle by which the D5 branes are rotated. 
If we add all the
above discussed terms and integrate out the adjoint field $\Phi$, 
the following superpotential is obtained:
\begin{equation}
\label{marginal}
\frac{g^2}{2 m} [\epsilon^{ij}\epsilon^{kl}\mbox{Tr} (A_i
B_k
A_j B_l) + \epsilon^{ij}\epsilon^{mn}\mbox{Tr} (C_i D_m C_j
D_n)]
\end{equation}
This is the term obtained by rotating all the branes by the same angle,
considering the terms which appear and then integrating out the massive
fields (only the adjoint field acquires mass).
Before rotation, the theory was ${\cal N} = 4$ and we could exchange the 
positions of the NS and D5 branes. After rotation, in order to be able to
be allowed to do that, all the branes need to be rotated by the same angle.
This implies that all the terms in (\ref{mass}) are needed in order to 
recover a self-dual theory. 
As explain in \cite{ms} we 
actually require that the ${\bf Z_2}$ symmetry exchanging the two ${\bf Z_2}$
orbifolds is preserve which determines a marginal deformation from one
self-dual model to another. The theory has a line of fixed points which
contains the ${\cal N} = 4$ theory. 

So, the field theory obtained by taking M theory on ${\bf S^7/ Z_2 \times
Z_2}$ 
is in our case equivalent to the ${\cal N} = 2$ three dimensional theory
obtained by adding the
superpotential (\ref{marginal}) to the $U(N) \times U(N)$ gauge theory
with
the matter content given by the fields A, B, C, D.

We want to make an important observation here. The brane configuration
described is invariant under an S-duality which replaces the NS branes by
D5 branes , the D5 branes by NS branes and leaves the D3 branes invariant.
This S-duality replaces the A and B fields by C and D fields and
vice-versa. In equation (\ref{super}), $C_i$ and $D_i$ appear in the first
term and $A_i$ and $B_i$ appear in the second term. Equation (\ref{mass})
remains invariant under exchanging of A, B and C, D fields. 
By remembering that the two $U(1)$ groups of the conifold act on pairs
$A, B$ and $C, D$, the S-duality just inverts this two $U(1)$ groups.
One of the $SU(2)$ groups will act upon $(A_1, A_2)$, one on
$(B_1, B_2)$ and the third one on $S' = C_2 D_1, T' = C_1 D_2$. 
\vskip .2cm

{\it Conifold Theory}

\vskip .2cm

We are discussing the field theory as obtained by starting from the
conifold theory.

The parameterization of the conifold is given in terms of the fields
$ S, T, C_{i} D_{i}$ as done in section 2. As done at the end of section 
2, we consider $S = A_2 B_1, T = A_1 B_2$. We consider a U(1)
gauge theory with ${\cal N} = 2$ in three dimensions and introduce 
$C_1, D_1$ as chiral multiplets with charge 1 and $C_2, D_2$
as chiral multiplets with charge -1. The fields $A_2, B_2$ are neutral under this
gauge group. We have equation (\ref{1}):
\begin{equation}
|B_1|^2 + |C_1|^2 + |C_2|^2 = |A_1|^2 + |D_1|^2 + |D_2|^2
\end{equation} 
But this is just the D auxiliary field
of the U(1) vector multiplet, by considering terms 
which do not involve the adjoint scalar obtained
after reduction from ${\cal N} = 1, D = 4$, so the moduli
space of vacua is part of the conifold. 
If we consider now another one  U(1)
gauge theory with ${\cal N} = 2$ in three dimensions and introduce
$B_2, C_1, D_1$ as chiral multiplets with charge 1 and $C_2, D_2, A_2$
with
charge -1. The fields $A_1, B_1$ are neutral under this
gauge group. We have equation (\ref{2}):
\begin{equation}
|B_2|^2 + |D_1|^2 + |D_2|^2 = |A_2|^2 + |C_1|^2 + |C_2|^2
\end{equation}
which can be again interpreted as a D auxiliary field
of the U(1) vector multiplet. 
In the paper \cite{kw}, one of the two gauge groups lives on the
worldvolume of their D3 branes so the conifold is identified with the 
moduli space of vacua of one of the $U(1)$ groups. Here we do not have any
$U(1)$ group on the two-branes worldvolume. So we need to identify the
moduli spaces of both $U(1)$ groups with the two branches of the 
conifold (\ref{1}),(\ref{2}). It is  now clear why do we need two
equations for the conifold as compared with \cite{kw} where there is only one.

The result is that we have a theory with the
gauge group $U(1) \times U(1)$ with chiral multiplets 
$A_1, B_1, A_2, B_2$ and $C_1, C_2$ and $D_1, D_2$ with charges 
(-1, 0), (1, 0), (0, -1), (0, 1), (1, -1) , (-1, 1) respectively. 
The chiral multiplets describe the M2 brane motion on the conifold.
The model can be considered to describe the low energy behavior of a 
threebrane on $M_3 \times Y$. 

If we have N M2 branes the gauge theory is generalized to
a $U(N) \times U(N)$ gauge theory with the same field content as for the
orbifold discussion. Now the chiral fields $C_i, D_i$ are matrices,
the chiral fields $A_i$ are  row vectors and $B_i$ are column vectors.
It is then natural to use the $S, T$ fields in order to have all the
fields given by matrices. If all the matrices are diagonal, the diagonal
entries of $C_i, D_i$ give us the positions of the N M2 branes at
distinct points on the conifold. The gauge group is broken to a product of 
U(1) factors. But there are the extra-diagonal entries which need to be
given masses in order to integrate out the unwanted massless chiral
multiplets and we are going to do it by introducing a superpotential
that does so. The superpotential should preserve 
the symmetry  of the conifold $Y$  i.e. $SU(2) \times
SU(2) \times SU(2) \times U(1)_{R}$ symmetry.
The $U(1)_{R}$ is the R-symmetry inherited by the ${\cal N} = 2$
three dimensional theory from its reduction from 4 dimensions. All the
fields $A_i, B_i, C_i, D_i$ have charge 1/2 and the fields $S, T$ have
charge 1. A superpotential $SU(2) \times
SU(2) \times SU(2)$ invariant and having $U(1)_R$ charge 2 can be written 
as: 
\begin{equation}
W = \lambda ( \mbox{Tr}(S T - T S) + \epsilon^{ij} \epsilon^{kl}
\mbox{Tr}(C_i D_k C_j D_l))
\end{equation}
where all the fields are now $N \times N$ matrices. If we now use the
definition of S and T , we can rewrite the superpotential as:
\begin{equation}
\label{marginal1}
\frac{g^2}{2 m} [\epsilon^{ij}\epsilon^{kl}\mbox{Tr} (A_i
B_k
A_j B_l) + \epsilon^{ij}\epsilon^{mn}\mbox{Tr} (C_i D_m C_j
D_n)]
\end{equation}
where the products $A_i B_k$ are to be understood as $N \times N$
matrices. This has charge 2 and is $SU(2) \times
SU(2) \times SU(2)$ invariant. The off diagonal components receive mass
from the superpotential (plus Higgs mechanism).  

We observe that the above superpotential is the same with the one
of (\ref{marginal}). This is a
marginal operator which takes us from a conformal theory to a
new conformal field theory.

We now have the ingredients to state the result of this paper:

{\it M theory on $AdS_4 \times Q^{1,1,1}$ is equivalent to the theory
obtained by starting with $U(N) \times U(N)$ theory with two copies of
$(N, \bar{N}) \oplus (\bar{N}, N)$ and four fields in $(N, 1), (\bar{N},
1), (1, N), (1, \bar{N})$ flowing to an infrared fixed point and then 
perturbed by the potential (\ref{marginal}).}

We end this section with a discussion over the conifold description of the
the configuration which is the S-dual of the one considered before.
By remembering that the two $U(1)$ groups of the conifold act on pairs
$A, B$ and $C, D$, the S-duality just inverts this two $U(1)$ groups.
One of the $SU(2)$ groups will act upon $(A_1, A_2)$, one on
$(B_1, B_2)$ and the third one on $S' = C_2 D_1, T' = C_1 D_2$.

As a conclusion, in this paper we extended the idea of comparing 
string/M theory at conifold singularities and superconformal field
theories to the case of $AdS_4 \times Q^{1,1,1}$ which gives superconformal
field theory on three dimensions. One important development is the one
also described in \cite{kw} i.e. the case of $AdS_4 \times V_{5,2}$
where $V_{5,2}$ is the seven dimensional Einstein homogeneous space
  SO(5)/SO(3)
obtained as a link (or horizon) of the 
singularity of a quadric hypersurface in ${\bf C}^5$.
 
\section{Acknowledgments}
We thank Donu Arapura and Prabhakar Rao for valuable discussions
on various parts of the section 2. 
We thank
Igor Klebanov and David Lowe for important comments on the manuscript. We
are grateful to 
Matthew Strassler for helping us at various points of this project, for
encouraging us and especially for sharing preliminary versions of his
work. We would like to especially thank him for suggesting to us that the 
${\bf C^2/Z_2 \times C^2/Z_2}$ is the proper model to consider because 
models with self mirror-symmetry have nice marginal operators and thus
being a crucial choice for our purposes. We
thank Changhyun Ahn for discussions at the initial stage of this
project.


\end{document}